\documentclass[aps,prl,twocolumn,amsmath,amssymb,showpacs]{revtex4}

\usepackage{amsmath}
\usepackage{graphicx}
\newcommand{\e}{{\rm e}}

\begin{document}

\title{Level statistics in arithmetical and pseudo-arithmetical chaos}
\author{
Petr Braun$^{1,2}$, Fritz Haake$^1$}

\address{$^1$Fachbereich Physik, Universit{\"a}t Duisburg-Essen,
  47048 Duisburg, Germany
  \\$^2$Department of Theoretical Physics, Saint-Petersburg University,
  198504 Saint-Petersburg, Russia}


\begin{abstract}

  We resolve a long-standing riddle in quantum chaos, posed by certain
  fully chaotic billiards with constant negative curvature whose
  periodic orbits are highly degenerate in length. Depending on the
  boundary conditions for the quantum wave functions, the energy
  spectra either have uncorrelated levels usually associated with
  classical integrability or conform to the ``universal'' Wigner-Dyson
  type although the classical dynamics in both cases is the same. The
  resolution turns out surprisingly simple. The Maslov indices of
  orbits within multiplets of degenerate length either yield equal
  phases for the respective Feynman amplitudes (and thus Poissonian
  level statistics) or give rise to amplitudes with uncorrelated
  phases (leading to Wigner-Dyson level correlations). The recent
  semiclassical explanation of spectral universality in quantum chaos
  is thus extended to the latter case of ``pseudo-arithmetical'' chaos.
\end{abstract}

\maketitle

{\it Introduction}: After more than two decades of investigations, the
famous BGS conjecture \cite{BGS} has recently found a semiclassical
explanation\cite{Berrydiag,Sieber,smalltau,taugt1,bible}: The quantum
level spectra of classically chaotic systems display fluctuations
conforming to the random-matrix-theory (RMT) predictions for the
Wigner-Dyson universality classes.

However, there is the notable exception of ``arithmetical'' systems
which are fully chaotic classically but display quantum spectral
statistics close to Poissonian, a behavior usually associated with
integrable classical motion
\cite{BolteSteiner,Bolte,Bogosteiner,Bogomolnygeorgeot,Bogomolnyleshouches,
  Marklof}. Quantum mechanically these exceptional dynamics exhibit an
infinite number of the so called Hecke operators commuting with
the Hamiltonian. Therefore the energy spectrum falls into
practically independent multiplets such that nearby levels bear no
correlation. On the classical side the periodic-orbit action
spectra of such systems are distinguished by a degeneracy
exponentially growing with the orbit period.

On the other hand, by merely changing the boundary conditions for
the quantum wave functions of some such exceptional arithmetical
systems one can  retrieve universal spectral fluctuations \`a la
Wigner and Dyson while not at all changing the classical dynamics.
It is customary to speak of pseudo-arithmetical systems then
\cite{earlySteiner,Steinersubtle,Ninnemann}.

The strikingly different quantum behavior of arithmetical and
pseudo-arithmetical systems might raise doubts about the validity of the
recent semiclassical explanation of universal quantum spectral
fluctuations under conditions of classical chaos; after all, the
classical dynamics are identical for the systems under discussion, and
so appear, on first sight, the Gutzwiller type semiclassical
periodic-orbit expansions. Various
suggestions were ventured for the effect of the boundary
conditions, among them a distinction between the orbit classes contributing
to the Selberg trace formula applicable (and exact) in the
arithmetical case, and the Gutzwiller formula applicable in the
pseudo-arithmetical case.

We show here that the explanation is much simpler and lies in
special properties of periodic orbits. Due to these
peculiarities all equal-length orbits (save for a negligible
fraction) of an arithmetical system contribute Feynman amplitudes
with the same Maslov phase; their constructive interference makes
for nonuniversal spectral statistics. In the pseudo-arithmetical
case these phases vary randomly within a degenerate-action
multiplet such that destructive interference makes the high action
degeneracy ineffective. The difference between the two cases is
most easily revealed for the diagonal approximation to the
spectral form factor, and therefore that approximation will play a
central role here; off-diagonal corrections will be discussed
briefly in the end.

{\it The billiard $T^*(2,3,8)$}: We shall not deal with the
alternative arithmetical/pseudo-arithmetical in full generality but
prefer to work with a representative example, the so called triangular
billiard $T^*(2,3,8)$.  That system was first considered in the
studies of the free motion on the surface of constant negative
curvature tesselated by regular
octagons\cite{Bohigastriangle}. Desymmetrization of the regular
octagon necessary to get rid of the rotational and reflection symmetry
leads to a triangular fundamental domain with the angles $\pi /2,\pi
/3,\pi /8$. Depicted inside the Poincar\'{e} disk $\left\vert
  z\right\vert \leq 1$ in the complex plane, the triangle has its
hypotenuse $\left( N\right) $ and the longer leg $\left( L\right)
$ directed along two diameters whereas the shorter leg $\left(
M\right) $ looks like an arc (Fig. 1). A classical periodic orbit
folded into the triangle looks like a sequence of arcs
mirror-reflected from the sides of the triangle. The classical
motion is completely chaotic, all periodic orbits having the same
Lyapunov constant.  The multiplicities in the length spectrum of
periodic orbits grow exponentially with the length, like $\exp
(l_{\gamma }/2)$, where $l_\gamma$ is made dimensionless by
referral to a scale fixed by setting the curvature to $-1$; since
action is proportional to the orbit length the action spectrum
also has an exponentially growing degeneracy.

\begin{figure}
\begin{center}
\includegraphics[scale=1.0] {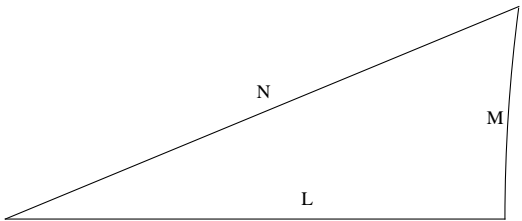}
%  {rs}
\end{center}
\caption{Fundamental domain of $T^*(2,3,8)$} \label{fig}
\end{figure}

The quantum energy levels for $T^*(2,3,8)$ are found as
eigenvalues of the Laplace-Beltrami operator. The boundary
conditions can be either Dirichlet or Neumann. There are $2^3=8$
quantum mechanical problems, all related to the same classical
system. In problems stemming from desymmetrization of the regular
octagon, the boundary conditions on the triangle sides are chosen
to obtain the spectrum for a particular irreducible
representation.  The boundary conditions on the hypotenuse $N$ and
the shorter leg $M$ must then be the same, i.e., both Dirichlet or
both Neumann; four such possibilities exist all of which lead to
arithmetical systems with near-Poissonian spectral statistics. The
remaining four choices where $N$ and $M$ host different boundary
conditions, lead to pseudo-arithmetical systems with the
Wigner-Dyson statistics of the orthogonal universality class.

{\it Form factor and diagonal approximation}: The
spectral form factor following from Gutzwiller`s trace formula is
a double sum over periodic orbits,
\begin{eqnarray}\nonumber
K\left( \tau \right) &\sim& \left\langle \sum_{\gamma ,\gamma
^{\prime }}A_{\gamma }A_{\gamma ^{\prime }}\exp \left[
i\frac{S_{\gamma }-S_{\gamma ^{\prime }}}{\hbar }-i\frac{\left(
\mu _{\gamma }-\mu _{\gamma ^{\prime }}\right) \pi }{2}\right]\right.\\
&&\hspace{1cm}\times \left.\delta \left( \tau T_{H}-\frac{T_{\gamma }+T_{\gamma
^{\prime }}}{2}\right) \right\rangle ,
\end{eqnarray}%
where $S_{\gamma },T_{\gamma ,}\mu _{\gamma },A_{\gamma
}=A_{\gamma }^*$ are action, period, Maslov index, and stability
coefficient of the orbit $\gamma $; the Heisenberg time
$T_{H}=\frac{2\pi
  \hbar }{\Delta }$, with $\Delta $ the mean level spacing, is used as
a unit of time such that $\tau$ becomes a dimensionless time; the
angular brackets $\left\langle \ldots \right\rangle $ denote averages
over the energy shell and over a small $\tau$ interval.

Of special interest are the pairs of orbits obviously immune against
destructive interference of their contributions, namely with the
same action and Maslov index. For generic chaotic dynamics these are
the trivial pairs $\gamma^{\prime }=\gamma $ and, if time reversal is
allowed, the pairs of mutually time reversed orbits, $\gamma ^{\prime
}=\gamma ^{\mathrm{TR}}$. Discarding all other pairs one gets Berry's
diagonal approximation \cite{Berrydiag},
 \begin{equation}\label{diagap}
 K_{\mathrm{diag}}\left( \tau \right) \sim
\sum_{\gamma }A_{\gamma }^{2}\delta \left( \tau T_{H}-T_{\gamma
}\right) =g\tau\,;
 \end{equation} here $g$ is the average
multiplicity of the action spectrum which is 1 in the absence of
time reversal (unitary universality class) or 2 in
the presence of time reversal (orthogonal class); the result
yields the first-order term of a power series in $\tau$, in
agreement with RMT \cite{smalltau}.

Turning to arithmetical chaos we can carry over the foregoing
reasoning, except that the multiplicity is now exponentially large,
$g\propto \e^{l/2}$.  However, orbits with the same length bear no
geometric similarity, and it is not at all obvious that their Maslov
phases are the same.

As will be shown, the Maslov phases do indeed coincide for all
orbits in a length multiplet, a negligible fraction apart, in the
arithmetical case. Consequently, the form factor exhibits almost instant
increase at $\tau\geq 0$, similar to the jump of the integrable case
\cite{bible}, $K(0)=0$ and $K(\tau)=1$ for $\tau>0$.

In contrast, the Maslov index of pseudo-arithmetical systems will
turn out to fluctuate randomly within each fixed-action multiplet.
The usual diagonal approximation with $g=2$ then holds since in
each multiplet only the pairs $\gamma'=\gamma$ and
$\gamma'=\gamma^{\rm TR}$ escape destructive interference. The
diagonal approximation thus suggests universal spectral
fluctuations.

The Maslov index of an orbit in our billiard is determined only by
the number $N_D$ of reflections from the sides with the Dirichlet
boundary condition: each such reflection changes the Maslov phase
by $\pi $. Therefore the contribution of an equal-action pair
$\gamma ,\gamma ^{\prime }$ is $ A_{\gamma }^{2}\left( -1\right)
^{N_{D}-N_{D}^{\prime }}$ where $ N_{D},N_{D}^{\prime }$
respectively refer to $\gamma $ and $\gamma ^{\prime }$; in long
orbits both $ N_{D}$ and $N_{D}^{\prime }$ are large pseudo-random
integers. We shall demonstrate that in arithmetical systems
equal-length orbits have, in their overwhelming majority, $N_{D}$
of the same parity (even or odd) such that contributions of all
pairs of them are positive and add up. On the contrary,
pseudo-arithmetical systems have uncorrelated parities of
$N_{D},N_{D}^{\prime }$ and the equal-action contributions
mutually cancel, apart from the standard pairs of the orthogonal
universality class.

{\it Numerical observations}: The triangular billiard affords symbolic
dynamics, and calculating its orbits involves generating all allowed
sequences of symbols $L,M,N$ each standing for the visit of the
respective side. We denote by $\lambda _{\gamma },\mu _{\gamma },\nu
_{\gamma }$ the number of symbols $L,M,N$ in an orbit $\gamma $; the
total number of symbols in $\gamma $ is $n_{\gamma }=\lambda _{\gamma
}+\mu _{\gamma }+\nu _{\gamma } $. Studying up to a million orbits we
find:

--- Orbits within a given length multiplet $\Lambda $ almost always
have $ n_{\gamma }$ with the same parity. We can therefore speak
about $\Lambda _{g} $- and $\Lambda _{u}$-multiplets depending on
parity of $n_{\gamma },\gamma \in \Lambda $. Exceptions are
extremely rare and in fact amount to a negligible fraction: E.g.,
among approximately $257000$ orbits with the length $l_{\gamma
}<16$ grouped into more than $13000$ length multiplets, only 4
multiplets are \textquotedblleft $gu$-degenerate\textquotedblright
, i.e., contain orbits with both even and odd number of symbols;
these offenders have lengths
$l=10.6999964,\;12.2422622,13.7571382,\;15.2857092$.

--- \textit{All} orbits in a given $\Lambda$ without
$gu$-degeneracy have $\lambda _{\gamma }$ of the same parity which
automatically leads to definite parity of $\mu _{\gamma }+\nu
_{\gamma }$. On the other hand, $\mu _{\gamma }$ and $\nu _{\gamma
}$ separately have no definite parity within $\Lambda $.

These observations, in particular the rarity of multiplets with
$gu$-degeneracy, suffice to explain the sensitivity of the level
statistics to the boundary conditions. We denote by $ \Phi _{L}$ the
phase jump on reflection from the side $L$, which is $0$ for the
Neumann and $\pi $ for the Dirichlet condition on $L$; similarly,
$\Phi _{M},\Phi _{N}$ denote the phase jumps on $M$ and $N$. The
contribution to the diagonal form factor of a pair of orbits $(\gamma
\gamma ^{\prime })$ belonging to the same $\Lambda $ is
\begin{eqnarray*}
K_{\gamma \gamma ^{\prime }} &=&A_{\Lambda }^{2}\e^{ i\left(
\lambda _{\gamma }-\lambda _{\gamma ^{\prime }}\right) \Phi
_{L}+i\left( \mu _{\gamma }-\mu _{\gamma ^{\prime }}\right) \Phi
_{M}+i\left( \nu _{\gamma
}-\nu _{\gamma ^{\prime }}\right) \Phi _{N}}  \\
&=&A_{\Lambda }^{2}\e^{ i\left( \mu _{\gamma }-\mu _{\gamma
^{\prime }}\right) \Phi _{M}+i\left( \nu _{\gamma }-\nu _{\gamma
^{\prime }}\right) \Phi _{N}} ;
\end{eqnarray*}%
the phase proportional to $\Phi _{L}$ disappears since, in the
absence of the $gu$-degeneracy, $\lambda _{\gamma }-\lambda
_{\gamma ^{\prime }}$ is even. If the boundary conditions on $M$
and $N$ are different (the pseudo-arithmetical case) one of $\Phi
_{M},\Phi _{N}$ is zero and another one $\pi $. The contribution
$K_{\gamma \gamma ^{\prime }}$ can then be of any sign and, summed
over all pairs of a large multiplet, except the standard $\gamma
^{\prime }=\gamma ,\gamma ^{\mathrm{TR}}$, averages to zero. The
form factor will be the Wigner-Dyson one for the orthogonal
universality class, $g=2$.

On the other hand, in  arithmetical systems with the same boundary
conditions on the sides $M,N$ we have $\Phi _{M}=\Phi _{N}\equiv
\Phi _{MN},$ and the contribution of every pair within $\Lambda$
will be positive since $\mu _{\gamma }+\nu _{\gamma }-\mu _{\gamma
^{\prime }}-\nu _{\gamma ^{\prime }}~$is even,
\begin{equation*}
K_{\gamma \gamma ^{\prime }}=A_{\Lambda }^{2}\e^{ i\left( \mu
_{\gamma }+\nu _{\gamma }-\mu _{\gamma ^{\prime }}-\nu _{\gamma
^{\prime }}\right) \Phi _{MN}} =A_{\Lambda }^{2}.
\end{equation*}%
Eq.(\ref{diagap}) then applies with the abnormally high value
of $g$ bringing about the nearly vertical rise of $K(\tau)$ at $\tau\geq 0$.

{\it Analytic reasoning}: Using group-theoretical properties
\cite{Ninnemann} of $T^*(2,3,8)$ we have substantiated these
results analytically. The symbols $L,M,N$ can be associated with
transformations mapping the interior of the Poincar\'{e} disk
$|z|\leq 1$ onto itself. Each elementary operation involves
complex conjugation $K:z\rightarrow z^{\ast }$ followed by a
M\"obius transformation $z\rightarrow (az+b)/(b^{\ast }z+a^{\ast
})$; the three respective matrices $\big({a \atop b^*}{b \atop
a^*}\big)$ can be chosen as
\begin{eqnarray}\label{matrix for symbol}\nonumber
\rho _{L} &=&
\begin{pmatrix}
1 & 0 \\
0 & 1
\end{pmatrix}
, \\
\rho _{M} &=&
\begin{pmatrix}
i\left( 1+\frac{\sqrt{2}}{2}\right) \gamma  & -i\frac{\sqrt{2}}{2}\beta  \\

i\frac{\sqrt{2}}{2}\beta  & -i\left( 1+\frac{\sqrt{2}}{2}\right) \gamma
\end{pmatrix}
, \\ \nonumber
\rho _{N} &=&
\begin{pmatrix}
\left( \frac{1+\sqrt{2}}{2}+\frac{i}{2}\right) \gamma  & 0 \\
0 & \left( \frac{1+\sqrt{2}}{2}-\frac{i}{2}\right) \gamma
\end{pmatrix}
\end{eqnarray}
where $\alpha ,\beta ,\gamma $ denote the quartic irrationalities
\begin{equation}
\alpha =\sqrt{\sqrt{2}-1},\quad \beta =\sqrt{\sqrt{2}},\quad \gamma =\sqrt{2-%
\sqrt{2}}\,.
\end{equation}

Let us write the code of the orbit $\gamma $ starting with an
arbitrary symbol and multiply the associated elementary operators.
The product is either a pure M\"obius transformation with a
certain matrix $\rho _{\gamma },$ if the number of symbols
$n_{\gamma }$ is even, or $K$ followed by $\rho _{\gamma }$ (odd
$n_{\gamma }$). The cumulative transformation leaves invariant a
circle in the complex plane (an invariant geodesic of the
transformation) crossing the fundamental domain; its part inside
the domain is the piece of $ \gamma $ between the two bounces
against the sides given by the first and last symbol of the code.
Cyclically shifting the code by one symbol one analogously gets
the next orbit piece,  and so forth \cite{Ninnemann}. The matrix
$\rho _{\gamma }$ yields the orbit length as
\begin{eqnarray}
2\cosh \frac{l_{\gamma }}{2} &=&\mathrm{Tr}\rho _{\gamma },\text{ \quad }%
n_{\gamma }\text{ even,}  \label{sinhcosh} \\
2\sinh \frac{l_{\gamma }}{2} &=&\mathrm{Atr}\,\rho _{\gamma },\quad \text{ }%
n_{\gamma }\text{ odd,}  \notag
\end{eqnarray}%
where $\mathrm{Atr}\,\rho $ is the sum of the off-diagonal
elements of $\rho .$

It can be shown by induction that the matrices $\rho _{\gamma }$
can be of two arithmetical types (\cite{Ninnemann}, Appendix B),
\begin{eqnarray}
\rho ^{(1)} &=&%
\begin{pmatrix}
\frac{1}{2}\left( u_{1,R}+iu_{1,I}\right)  & \frac{1}{2}\left(
v_{1,R}+iv_{1,I}\right) \alpha  \\
\frac{1}{2}\left( v_{1,R}-iv_{1,I}\right) \alpha  & \frac{1}{2}\left(
u_{1,R}-iu_{1,I}\right)
\end{pmatrix}%
,  \label{twotypes} \\
\rho^{(2)} &=&%
\begin{pmatrix}
\frac{1}{2}\left( u_{2,R}+iu_{2,I}\right) \gamma  & \frac{1}{2}\left(
v_{2,R}+iv_{2,I}\right) \beta  \\
\frac{1}{2}\left( v_{2,R}-iv_{2,I}\right) \beta  & \frac{1}{2}\left(
u_{2,R}-iu_{2,I}\right) \gamma
\end{pmatrix}%
.  \notag
\end{eqnarray}%
Here $u_{k,R},v_{k,R},u_{k,I},v_{k,I}$ are algebraic integers,
\begin{eqnarray}\label{algebraicinteger}
u_{k,R} &=&m_{k,R}+n_{k,R}\sqrt{2}, \\
v_{k,R} &=&p_{k,R}+q_{k,R}\sqrt{2},\quad k=1,2,\nonumber
\end{eqnarray}%
with integers $m_{k,R},n_{k,R},p_{k,R},q_{k,R}$; the imaginary
parts have the same appearance. Appending  a symbol $L$ (pure
complex conjugation) to the code of the orbit doesn't change the
type of $\rho _{\gamma }$ whereas appending $M$ or $N$ toggles the
type, $\rho^{(1)}\leftrightarrow\rho^{(2)}$. The matrices $\rho
_{M}$ and $\rho _{N}$ belong to the type $\rho^{(2)}$; therefore
$\rho _{\gamma }$ belongs to the type $\rho^{(1)}$ if the sum
$\mu_{\gamma }+\nu _{\gamma }$ of the number of symbols $M,N$ in
the code of $ \gamma $ is even, and $\rho _{2}$ if $\mu _{\gamma
}+\nu _{\gamma }$ is odd.

Eqs.~(\ref{sinhcosh},\ref{twotypes}) entail \textit{four} types of
orbit lengths:  Orbits with even numbers of symbols $n_{\gamma }$
have lengths
\begin{eqnarray}
\text{(a)}\quad 2\cosh \frac{l_{g}}{2}=u_{1,R}\,,\qquad
\text{(b)}\quad 2\cosh \frac{l_{g}}{2}=\gamma \,u_{2,R}\,,
\label{fourtypesab}
\end{eqnarray}
while orbits with odd  $n_{\gamma }$ have lengths
\begin{eqnarray}
\text{(c)}\quad 2\sinh \frac{l_{u}}{2}=\alpha
\,v_{1,R}\,,\qquad%\nonumber\\
\text{(d)}\quad 2\sinh \frac{l_{u}}{2}=\beta \,v_{2,R}\,.
\label{fourtypescd}
\end{eqnarray}
The integer components of $u,v$ in these equations are restricted
\cite{Ninnemann} by the inequalities
\begin{eqnarray}\label{inequalities}\nonumber
\mbox{(a)} \quad |m-n\sqrt{2}|<2\,,&&
\mbox{(b)} \quad |m-n\sqrt{2}|<\sqrt{2}\gamma\,,
\\
\mbox{(c)} \quad |p-q\sqrt{2}|<2\alpha\,,&&
\mbox{(d)} \quad |p-q\sqrt{2}|<\sqrt{2}\beta\,.
\end{eqnarray}

The length types (a,c) are connected with the matrices $\rho
_{\gamma }$ of the type $ \rho^{(1)}$ whereas (b,d) are connected
with $\rho^{(2)}$. Considering the connection between the type of $\rho
_{\gamma }$ and the code of $\gamma $ we see that the type of the
orbit length is uniquely defined by parities of the number of
bounces $\lambda _{\gamma }$ and $\mu _{\gamma }+\nu _{\gamma } $,
see Table~\ref{tab1}.

\begin{table}[h]
\caption{Orbit length types for different  parities of symbol
numbers} \label{tab1}
\begin{tabular}{ccc}
& $\lambda $ even & $\lambda $ odd \\
\hline
$\left( \mu +\nu \right) $ even & a & c \\
$\left( \mu +\nu \right) $ odd & d & b%
\end{tabular}
\end{table}

It is elementary to prove that the numbers in r.h.s. of the
equations (a),(b) in (\ref{fourtypesab}) cannot be equal unless
they are zero; the same is true for (c),(d) in
(\ref{fourtypescd}). Therefore a length multiplet can never
simultaneously contain orbits of the types (a) and (b) or (c) and
(d).  On the other hand, an orbit with an even number of symbols
$n_\gamma$ can have the same length as one with odd $n_\gamma$,
and such equality happens for some rare combinations of $
u_{k,R},v_{k,R}$; these are the cases of $gu$-degeneracy mentioned
above. Equating one of $l_a,l_b$ to one of $l_c,l_d$ we obtain an
equation for $u,v$ equivalent to two diophantine equations for the
integers $m,n,p,q$. (Not all solutions of these equations
correspond to really existing length multiplets since
(\ref{fourtypesab}), (\ref{fourtypescd})
 are only necessary conditions.) A careful
analysis of the latter equations (see below for details) for the
cases $l_a=l_c$ or $l_b=l_d$ reveals that the solutions form an
equidistant sequence $l^{(1)}_k=s^{I}k,\quad k=1,2,\ldots$ with
$s^{I}=2\,\mathrm {arsinh}\,2^{-1/4}$. Similarly solutions for
$l_a=l_d$ or $l_b=l_c$ are described by $l^{(2)}_k=s^{II}k,\quad
k=1,2,\ldots,$ with $s^{II}=2\,\mathrm
{arsinh}\,\sqrt{1+\sqrt{2}}$. E.g., the exceptional length
$l=10.6999964=7s^{I}$ pertains to two multiplets, (b) with
$m=138,n=97,$ and (d) with $p=88,q=63$.

 Since the exceptional lengths appear in equidistant sequences the
multiplets with $gu$-degeneracy are exponentially outnumbered, as the
length $l$ grows, by the  multiplets of definite type and
hence parity of $\lambda$ and $\mu+\nu$. The $gu$-degenerate
multiplets can thus be discarded for lengths corresponding to a
finite fraction of $T_H$.

To conclude, we have shown that ``all'' (all save for a negligible
fraction of multiplets) orbits with the same length of the $T^{\ast
}(2,3,8)$ billiard have the same parity of the number of bounces
against the longer leg $L$ of the triangle; this is also true for the
parity of the sum of the number of bounces against the sides $M,N$. As
a result ``all'' orbits with the same length in the arithmetical case
have the same Maslov index such that all diagonal terms in the form
factor are positive.  A similar mechanism must exist in all other
arithmetical systems with non-vanishing Maslov phases.  In
pseudo-arithmetical systems the Maslov phases of degenerate orbits are
uncorrelated.

{\it Off-diagonal corrections}:  We based our reasoning on the
diagonal approximation. But the essence of our
argument carries over to the higher-order terms of the
$\tau$-expansion valid for $\tau<1$ \cite{smalltau} as well as for
the behavior at times exceeding the Heisenberg time \cite{taugt1},
as far as the pseudo-arithmetical case is concerned: The high
multiplicity of length multiplets of orbits is rendered irrelevant
by destructive interference of random Maslov phases.

For arithmetical systems the interplay of high length degeneracy and
bunches of orbits that very nearly coincide in configuration space,
apart from reconnections in close self-encounters, may be more difficult
to capture. On the other hand, the Hecke symmetries intuitively
suggest (near) Poissonian level statistics.

{\it Technical note on exceptional multiplets}: We briefly
indicate how existence and uniqueness of the exceptional lengths
can be ascertained, for the four possible cases $l_a=l_c, l_a=l_d,
l_b=l_c, l_b=l_d$ which we shall refer to as $ac, ad, bc, bd$.
Starting with the case $ac$ we infer from
Eqs.~(\ref{fourtypesab},\ref{fourtypescd}) that the exceptional
lengths obey
\begin{equation} e^{\frac{l}{2}}=\frac{u+\alpha v}{2},\quad e^{ -\frac{l}{2}} =
\frac{u-\alpha v}{2}\,.
\end{equation}
Now take two solutions $u_{0}+\alpha v_{0}$ and $u+\alpha v$
corresponding to the lengths $l_{0},l$. Then since $\exp \left[ \pm \left(
\frac{l_{0}}{2}+\frac{l}{2}\right) \right] =\exp \left( \pm \frac{l_{0}}{2}%
\right) \exp \left( \pm \frac{l}{2}\right) $, the product
\begin{equation}
u^{\prime }+\alpha v^{\prime }=\frac{u_{0}+\alpha v_{0}}{2}\frac{u+\alpha v}{%
2}  \label{recur}
\end{equation}%
is a solution with length $ l^{\prime }=l+l_{0}$, provided
the integers in the rhs are divisible by 2.

The numerically found solution of smallest length  $
l_{0}^{I}=2s^{I}=3.05714183896200$ was
$u_{0}=2+2\sqrt{2},v_{0}=4+2\sqrt{2},$ or
$m_{0}=2,n_{0}=2,p_{0}=4,q_{0}=2$. Substituting the latter into
(\ref{recur}) we obtain
\begin{eqnarray}
m^{\prime } &=&m+2n+2q,  \label{direct} \\
n^{\prime } &=&m+n+p,  \nonumber \\
p^{\prime } &=&2m+2n+p+2q,  \nonumber \\
q^{\prime } &=&m+2n+p+q;  \nonumber
\end{eqnarray}%
obviously if $m,n,p,q$ are even then so are the primed numbers. We
thus face the sequence of solutions $ \left( \frac{u_{0}+\alpha
 v_{0}}{2}\right) ^{k}$ with the equidistant lengths $ kl_{0}^{I}$.

The transformation (\ref{direct}) can be inverted; the doubly primed
numbers
\begin{eqnarray}
m^{\prime \prime } &=&m+2n-2q,  \label{inverse} \\
n^{\prime \prime } &=&m+n-p,  \nonumber \\
p^{\prime \prime } &=&-2m-2n+p+2q,  \nonumber \\
q^{\prime \prime } &=&-m-2n+p+q;  \nonumber
\end{eqnarray}%
yield an orbit of length $l^{\prime \prime }=l-l_{0}^{I}$ such that a
ladder of decreasing lengths is obtained.

In order to make sure that the transformations
(\ref{direct},\ref{inverse}) really yield orbits we must check
that the inequalities (a,c) in (\ref{inequalities}) are not
violated.  To that end we note that the variables $x\equiv
(m-n\sqrt{2})/2,\; y\equiv (p-q\sqrt{2})/2\alpha$ span an
invariant subspace of the transformations
(\ref{direct},\ref{inverse}); the ensuing transformation
$(x,y)\to (x',y')$ is a two dimensional rotation preserving
$x^2+y^2$. Since the aforementioned minimal-length solution
$m_{0}=2,n_{0}=2,p_{0}=4,q_{0}=2$ has the property $x^2+y^2=1$ and
since that ``normalization'' is preserved we indeed conclude the
preservation of the inequalities under study.

Finally, let us demonstrate that there are no lengths of the type
$ac$ outside the equidistant-length ladder just established.
Momentarily assuming the existence of such a freak length we can
apply the length reducing transformation (\ref{inverse}) until
arriving at a reduced length $l''\in[0,l_0^I]$ which must obey
\begin{eqnarray}
1 &<&\cosh \frac{l^{\prime \prime }}{2}=\frac{m^{\prime \prime }+n^{\prime \prime
}\sqrt{2}}{2}<1+\sqrt{2}=\cosh \frac{l_{0}^{I}}{2}\,,\quad  \\ \nonumber
0 &<&\sinh \frac{l^{\prime \prime }}{2}=\left( p^{\prime \prime }+\sqrt{2}
q^{\prime \prime }\right) \frac{\alpha}{2} <\left( 2+\sqrt{2}\right) \alpha =\sinh
\frac{l_{0}^{I}}{2}\,.
\end{eqnarray}
On the other hand, the conservation of $x^{2}+y^{2}$ subjects the
freak to $\left( x^{\prime \prime }\right) ^{2}+\left( y^{\prime
    \prime }\right) ^{2}=\left( m^{\prime \prime }-n^{\prime \prime
  }\sqrt{2}%
\right) ^{2}/4+\left( p^{\prime \prime }-q^{\prime \prime
  }\sqrt{2}\right) ^{2}/4\alpha ^{2}=x^{2}+y^{2}\leq 2,$ and thus to
$\left\vert m^{\prime \prime }-n^{\prime \prime }\sqrt{2}\right\vert
<2\sqrt{2},\quad \left\vert p^{\prime \prime }-q^{\prime \prime
  }\sqrt{2}\right\vert <2\sqrt{2}\alpha$.  Upon checking the remaining
finite number of quadruplets of integers $m'',n'',p'',q''$ with lengths
inside the interval $ 0<l''<l_{0}^{I}$ we conclude that no freak can
exist.

The discussion of case $ad$ proceeds in close analogy with the
previously treated $ac$. The only differences are the replacements
of (i) the quartic irrationality $\alpha$ by $\beta$ and (ii) the
minimal length $l_0^I$ by $l_0^{II}=2s^{II}=4.8969$, the latter
corresponding to $u_{0}=6+4\sqrt{2},\quad v_{0}=4+4\sqrt{2}$. All
other lengths are given by $kl_{0}^{II},\quad k=1,2,\ldots.$ The
proof repeats the previous case word for word.

To treat case $bd$ we employ $\gamma =\alpha \beta $ to write the equations for the
lengths as
\begin{equation}
e^{\frac{l}{2}}=\frac{\beta \left( \alpha u+v\right) }{2},\quad e^{-\frac{l}{%
2}}=\frac{\beta \left( \alpha u-v\right) }{2}\,.  \label{bdexpexp}
\end{equation}
The smallest length corresponds to
$u_{m}=2+\sqrt{2},v_{m}=\sqrt{2}$ and is exactly $s^{I}$, i.e.,
half of the smallest length of case $ac$. Solutions of
(\ref{bdexpexp}) do not have in general the group property of the
case $ac$; in particular, if $l$ is an exceptional length of the
type $bd $ then $2l$ is not. Indeed, squaring the rhs of
(\ref{bdexpexp}) we get $\sqrt{2}\left( \alpha u\pm v\right)
^{2}/4$ which cannot be an algebraic number of the type $\beta
\left( \alpha u+v\right) /2$. On the other hand, all odd multiples
of $l$ do belong to the admissible type. The same reasoning as in
the cases $ac,ad$ shows that all solutions can be represented as $
\left( k+1/2\right) l_{0}^{I},$ $k=0,1,\ldots$.

The remaining case $bc$ is related to $bd$ by the replacements
$\alpha\leftrightarrow\beta$ and $l_{0}^I\to l_{0}^{II}$. All
solutions can be written as $\left( k+1/2\right)
l_{0}^{II},\,k=0,1,\ldots $.

We thank Eugene Bogomolny for useful discussions. Support by the
Sonderforschungsbereich SFB/TR12 of the Deutsche
Forschungsgemeinschaft is gratefully acknowledged.

\end{document}